\documentclass{emulateapj}
\lefthead{Laha et al.}

\usepackage{times}
\usepackage{url}
\usepackage{graphicx}
\usepackage[usenames]{color}
\DeclareGraphicsExtensions{.pdf,.png,.jpg,.mps,.eps,.ps}
\usepackage{amsmath}
\usepackage{natbib}
\usepackage{amsmath}
\usepackage{relsize}
\usepackage{pdflscape}
\usepackage{lscape}
\usepackage{rotating,booktabs}

\def\unit #1{\,{\rm #1}}

\newcommand\cmcubei{\rm \,\unit{cm^{-3}}}

\newcommand\fe{Fe\,{\sc iii}\,}

\def\aap{A\&A} \def\apjl{ApJ}  
 \def\apjs{ApJS}

\usepackage{graphicx}

\begin{document}

\title{Emission line ratios of Fe III as astrophysical plasma diagnostics.}

\author{Sibasish Laha\altaffilmark{1}, Niall B. Tyndall\altaffilmark{1}, Francis P. Keenan\altaffilmark{2}, Connor P. Ballance\altaffilmark{1}, \\ Catherine A. Ramsbottom\altaffilmark{1}, Gary J. Ferland\altaffilmark{3}, and Alan Hibbert\altaffilmark{1}.}\altaffiltext{1}{{Centre for Theoretical Atomic, Molecular and Optical Physics, School of Mathematics and Physics, Queen's University Belfast, Belfast BT7 1NN, Northern Ireland, U.K.}; {\tt email: s.laha@qub.ac.uk, sib.laha@gmail.com}}\altaffiltext{2} {Astrophysics Research Centre, School of Mathematics and Physics, Queen's University Belfast, Belfast BT7 1NN, Northern Ireland, U.K.}\altaffiltext{3} {Department of Physics and Astronomy, The University of Kentucky, Lexington, KY 40506, U.S.A;{\tt email: gary@g.uky.edu}}

%\pagerange{\pageref{firstpage}--\pageref{lastpage}} \pubyear{2013}

% \def\LaTeX{L\kern-.36em\raise.3ex\hbox{a}\kern-.15em
% T\kern-.1667em\lower.7ex\hbox{E}\kern-.125emX}

% \newtheorem{theorem}{Theorem}[section]

% \label{firstpage}

\begin{abstract}

Recent state-of-the-art calculations of A-values and electron impact excitation rates for \fe are used in conjunction with the Cloudy modeling code to derive emission line intensity ratios for optical transitions among the fine-structure levels of the 3d$^6$ configuration. A comparison of these with high resolution, high signal-to-noise spectra of gaseous nebulae reveals that previous discrepancies found between theory and observation are not fully resolved by the latest atomic data. Blending is ruled out as a likely cause of the discrepancies, because temperature- and density-independent 
ratios (arising from lines with common upper levels) match well with those predicted by theory. For
a typical  nebular plasma with electron temperature $T_{\rm e} = 9000$ K and electron density $N_{\rm e}=10^4 \cmcubei$, cascading of electrons from the levels $\rm ^3G_5$, $\rm ^3G_4$ and $\rm ^3G_3$ plays an important role in determining the populations of lower levels, such as 
$\rm ^3F_4$, which provide the density diagnostic emission lines of \fe, such as $\rm ^5D_4$ - $\rm ^3F_4$ at 4658\,\AA.  Hence further work on the A-values for these transitions is recommended, ideally including measurements if possible. However, some \fe ratios do provide reliable $N_{\rm e}$-diagnostics, such as 4986/4658. The \fe cooling function calculated with Cloudy using the most recent 
atomic data is found to be significantly greater at $T_e$ $\simeq$ 30000\,K than predicted with the existing Cloudy model. This is due 
to the presence of additional emission lines with the new data, particularly in the 1000--4000\,\AA\ wavelength region.

\end{abstract}


\vspace{0.5cm}

\section{Introduction.}

Emission lines arising from transitions among the fine-structure levels of the $\rm 3d^6$ configuration of \fe are widely 
observed in the optical spectra of astrophysical sources, including planetary nebulae, H~II regions and quasars  \citep[see, for 
example,][and references therein]{1978ApJ...222..384G,1993ApJ...410..430K,2003A&A...401.1119R,2009MNRAS.395..855M}.  These \fe 
transitions are also important tracers of Fe abundance in the case of H II regions and lowly ionised planetary nebulae, where they are often the only ionisation state of Fe detected in the optical band.

\citet{1978ApJ...222..384G} first noted the diagnostic potential of optical \fe lines, and subsequently several authors have generated theoretical electron temperature ($T_{\rm e}$) and density ($N_{\rm e}$) dependent emission line intensity ratios for this ion, and used these to 
determine plasma parameters for nebular sources \citep[for example,][and references therein.]{1993ApJ...410..430K,2001PNAS...98.9476K,2010ApJ...718L.189B}. However, there are longstanding inconsistencies between electron densities derived from 
different \fe ratios, as well as discrepancies with values of $N_{\rm e}$ determined from other species such as {O~{\sc ii}}, {S~{\sc ii}} and {Cl~{\sc iii}} with similar ionisation potentials to \fe and which hence should originate in nearby regions of the nebular plasma. For example, \citet{2011MNRAS.415..181F} in their study of the Saturn nebula NGC 7009 found electron densities in the range
$N_{\rm e}$ = 10$^{4.4}$--10$^{5}$ cm$^{-3}$ from 
several \fe ratios, more than an order of magnitude greater than those from {S~{\sc ii}} or {Cl~{\sc iii}}. Similarly, \citet{2003A&A...401.1119R}, in their study of the emission-line spectrum of the hot post-Asymptotic Giant Branch star HD~341617,
found that although most \fe line ratios indicate $N_{\rm e} \sim 10^4 \cmcubei$ (consistent with those from {O \sc{ii}}), 
several implied $N_{\rm e} \ge 10^5 \cmcubei$. 

Recently, \citet{2014ApJ...785...99B} have produced new state-of-the-art excitation rate data for \fe using the R-matrix
suite of packages, while \citet{2009ADNDT..95..184D} has previously calculated A-values for this ion using the highly sophisticated
CIV3 code \citep{1975CoPhC...9..141H,1991CoPhC..64..455H}. In this paper we use these data to generate \fe line intensity ratios, which we compare with both other theoretical results and with high spectral resolution observations, to investigate if the longstanding problems with this ion in nebular spectra 
can be resolved. The paper is arranged as follows. In Section 2 we discuss representative 
high resolution optical observations of \fe emission lines in nebular sources, while 
Section 3 contains details of the line ratio calculations. Results are presented and discussed in 
Section 4, and conclusions in Section 5.

%%%%%%%%%%%%%      Section 2

\section{Observations}\label{subsec:obs}

The \fe diagnostic emission lines in the optical region lie between $\sim$ 4000--5500 $\rm \AA$, and arise due to 
$\rm ^5D - {^3P}$, $\rm ^5D - {^3F}$, $\rm ^5D - {^3G}$ and $\rm ^5D - {^3H}$ transitions among levels of 
the $\rm 3d^6$ configuration. These are listed in Table 1. The \fe lines are in a crowded region of the spectrum,
 leading to the possibility of blending. Examples of close emission line pairs include: He {\sc{i}} 4009.25 \AA\ and \fe 
 4008.36 \AA; O {\sc{ii}} 4661.63 \AA\ and \fe  4658.05 \AA; [Fe {\sc{ii}}]  5273.35 \AA\ and \fe  
 5270.40 \AA\ \citep[see for example,][]{2002A&A...389..556R,2000ApJS..129..229B}. Hence, for the purpose of comparison between theory and observation in our study, we select only those observations which employ high resolution ($R \sim 10,000$) spectra, to
 ensure as far as possible that the \fe lines are not blended. We also focus on observations which have high signal-to-noise (S/N) ratios to facilitate the reliable detection of these weak lines. 

Our observational datasets consist of \fe line intensity ratios for (i) the hot post-Asymptotic Giant Branch star HD 341617, obtained by \citet{2003A&A...401.1119R} using the Keck telescope; (ii) the brightest knot of the Herbig Haro object HH~202 in the Orion nebula, studied by \citet{2009MNRAS.395..855M} using the Very Large Telescope at the European Southern Observatory; (iii) the Orion nebula H II region 
 by \citet{1998MNRAS.295..401E}, which employ data from the 2.1 m telescope at the Observatorio Astronomico Nacional (OAN) in Mexico;
 (iv) the Orion nebula H II region, this time obtained by \citet{2000ApJS..129..229B} using the 4 m telescope at the Cerro Tololo Inter-American Observatory (CTIO). Details of the observations may be found in the above references. In Table \ref{Table:obs1} the observed \fe line ratios from these datasets which are
density sensitive are summarised, while in Table \ref{Table:obs2} we list those arising from common upper levels and hence should be 
independent of $T_{\rm e}$ and $N_{\rm e}$. The errors in the line ratios from \citet{2003A&A...401.1119R} are assumed to be 10\%, as these authors 
do not quote any uncertainty estimates for their data. {This assumption is based on the fact that the Ryans et al. 
spectra are better in both spectral resolution and S/N than the Orion data of \citet{1998MNRAS.295..401E}, where they claim intensity errors of $<$\,10\%\ for lines of similar strength to those of \fe, yielding line intensity ratios with errors of $<$\,14\%. Hence adopting a 
10\%\ error for the HD 341617 data is probably an overestimate.}  
Note that all ratios are in energy units.

In Table \ref{Table:obs3} we list average values of electron temperature and density derived for our nebular sample
in the relevant references listed above. These plasma parameters were obtained using diagnostic line ratios in ions which have
similar ionisation potentials to that of \fe (30.7 eV) and hence should be emitted from a co-spatial region, and include
for example O {\sc{ii}} (35.1 eV), Cl {\sc{iii}} (39.6 eV) and N {\sc{ii}} (29.60 eV). 
\citet{2003A&A...401.1119R} could not estimate the temperature of the nebular plasma in HD 341617, due to the 
lack of reliable diagnostics, and hence adopted a value of 10,000 K from \citet{2000A&A...357..241P}. However, we note that most of the \fe line ratios are not particularly sensitive to $T_{\rm e}$, as discussed in Section 3.

%%%%%%%%%%%%%%     Section 3

\section{Theoretical line ratios.}\label{section:theory}

The Cloudy modeling code \citep{1998PASP..110..761F,2013RMxAA..49..137F} and CHIANTI suite of packages \citep{1997A&AS..125..149D,2015A&A...582A..56D} are employed to calculate \fe line intensity ratios. We have used several atomic datasets, including 
electron impact excitation rates (ECS) and transition probabilities (A-values) from \citet{2014ApJ...785...99B}. 
{These authors have calculated ECS using three methods, namely (i) intermediate coupling frame transformation (ICFT), (ii) 
Breit-Pauli R-matrix (BPRM) and (iii) Dirac Atomic R-matrix (DARC), each for the lowest 322 fine-structure levels. They found excellent agreement among all three calculations, and here we have used ECS values from the ICFT method, although we note that adoption of either of the other two \fe datasets leads to the same results.} 
Transition probabilities for $\sim 9000$ transitions among the lowest $285$ levels have also been taken 
from \citet{2009ADNDT..95..184D}, calculated with the 
general configuration interaction code (CIV3) and a large configuration set. There are significant differences in the A-values between the Badnell \& Ballance and Deb \& Hibbert studies. However, considering the more rigorous calculations by the latter 
with many configurations, we have adopted these in the final dataset. We employ the measured energies from NIST
for the 322 levels of Badnell \& Ballance, and correct their A-values for the energy differences between theory and experiment. 

Using the above atomic data, we have generated three different Cloudy models, termed CLOUDY1, CLOUDY2 and CLOUDY3. In the CLOUDY1 
model, the energy levels, A-values and ECS  are from \citet{2014ApJ...785...99B}, with a total of 51,681 transitions among 322 
levels. The energy level values from \citet{2014ApJ...785...99B} are consistently higher than those measured by NIST and are not ordered as per increasing NIST energies. The CLOUDY2 model comprises energies for the 322 levels from NIST, and energy-corrected A-values and ECS from \citet{2014ApJ...785...99B}. This is also the same set of data adopted in CHIANTI. The CLOUDY3 model is the same as CLOUDY2, except for the A-values of the $\sim 9000$ transitions among the lowest $285$ levels, which are from  \citet{2009ADNDT..95..184D}. Henceforth CLOUDY3 will be referred to as the `final' model. 

In addition to the above, we also consider a CLOUDY4 model that employs the \fe atomic data currently in Cloudy \citep{1996A&AS..119..523Z}, and are summarised in \citet{2013MNRAS.429.3133L}. As iron is one of the main elements responsible for maintaining thermal equilibrium in a nebular
plasma, we include CLOUDY4 in our study to assess any differences in plasma cooling rates when Cloudy is updated with the new \fe atomic data. We discuss this in detail in the Section 4.2.

\begin{figure*}
  \centering

\includegraphics[width=10cm,angle=0]{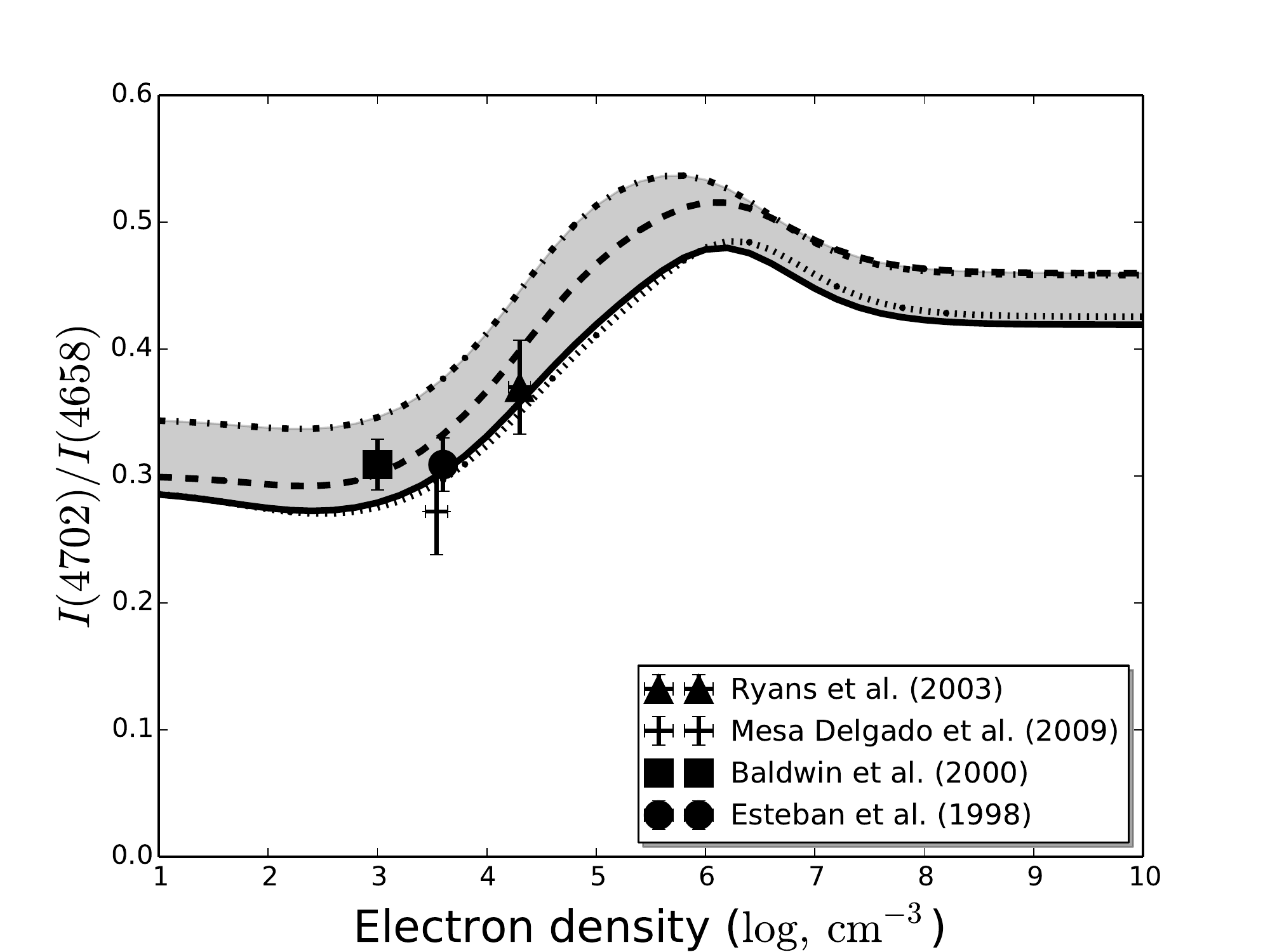}

\caption{Theoretical \fe line ratio 4702/4658 plotted as a function of electron density at an electron temperature $T_{\rm e}=9000$ K. The top curve is obtained using the CLOUDY2 model, the middle curve is with CLOUDY1 and the bottom with CLOUDY3. Observed data points, from the references listed in Section 2, are plotted at the values of density found for these sources from other diagnostic line ratios (see Section 2 for details). The grey band denotes the `error' in the theoretical line ratioS due to 
the different A-values adopted. The dotted line is the \fe line ratio with the CLOUDY3 model at $T_{\rm e}=15000$ K. }\label{fig:1}
\end{figure*}

\begin{figure*}
  \centering

\includegraphics[width=10cm,angle=0]{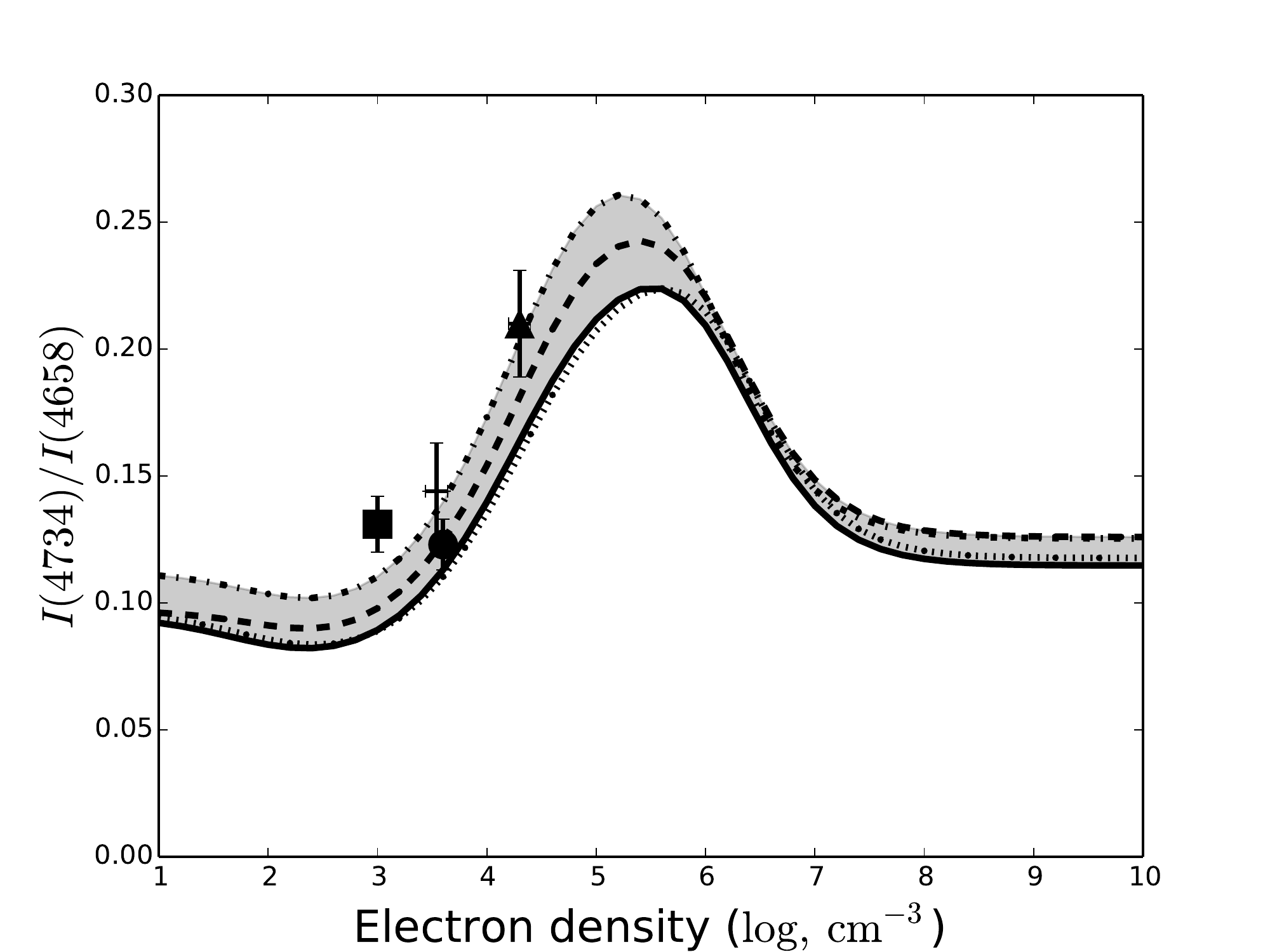}

\caption{Same as Figure \ref{fig:1} except for the ratio 4734/4658.}\label{fig:2}
\end{figure*}

\begin{figure*}
  \centering

\includegraphics[width=10cm,angle=0]{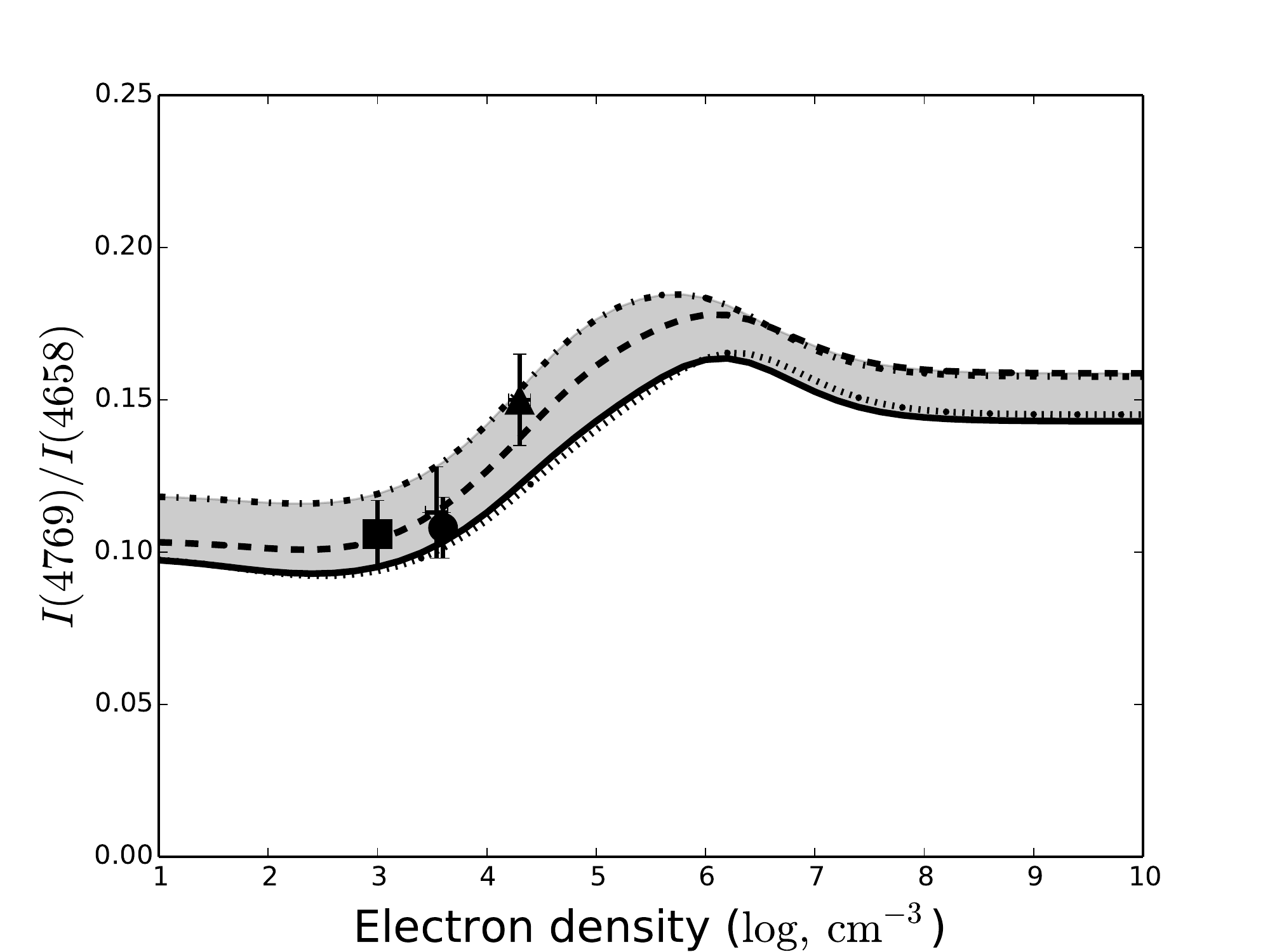}

\caption{Same as Figure \ref{fig:1} except for the ratio 4769/4658. }\label{fig:3}
\end{figure*}

\begin{figure*}
  \centering

\includegraphics[width=10cm,angle=0]{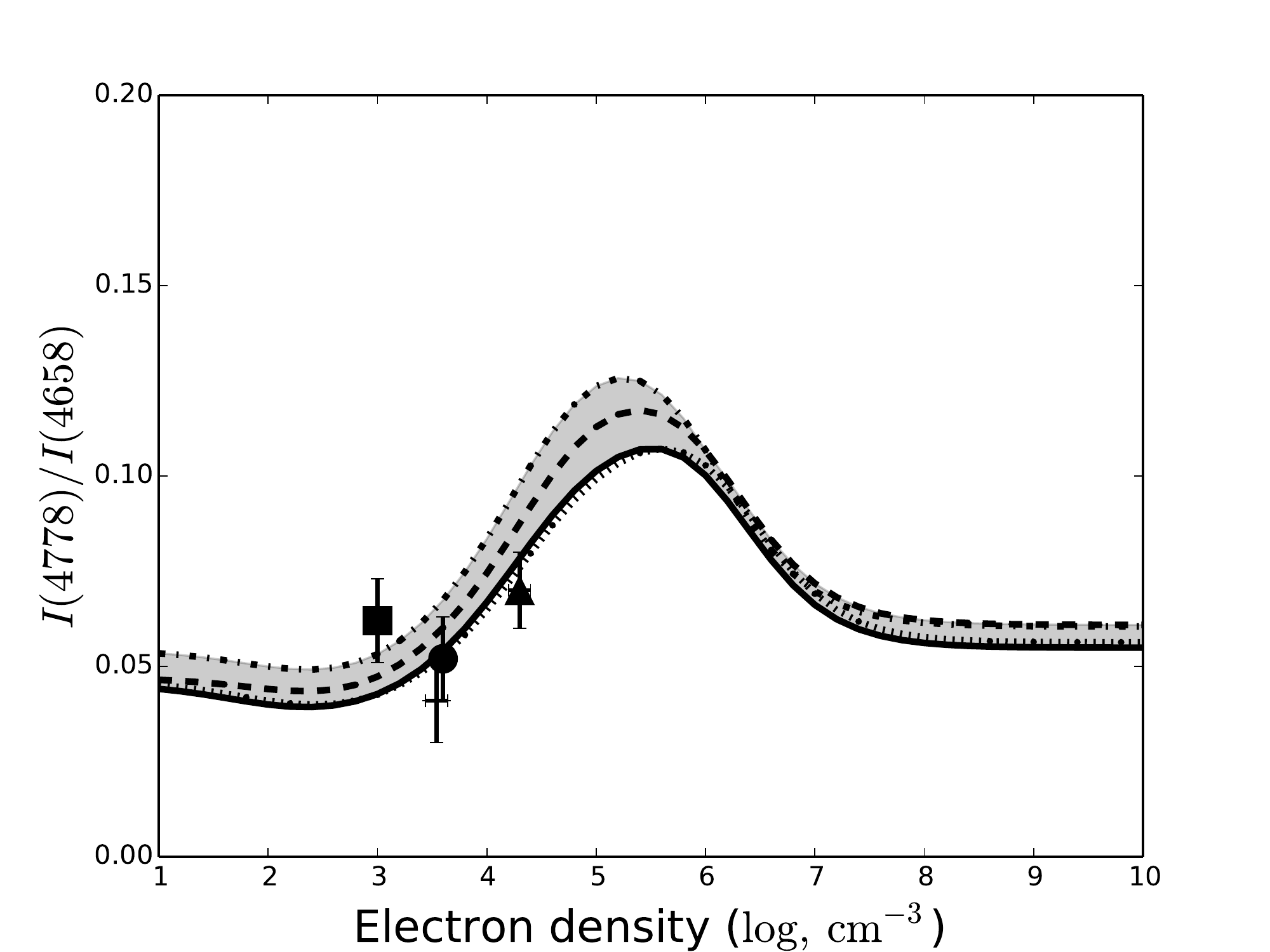}

\caption{Same as Figure \ref{fig:1} except for the ratio 4778/4658. }\label{fig:4}
\end{figure*}

\begin{figure*}
  \centering

\includegraphics[width=10cm,angle=0]{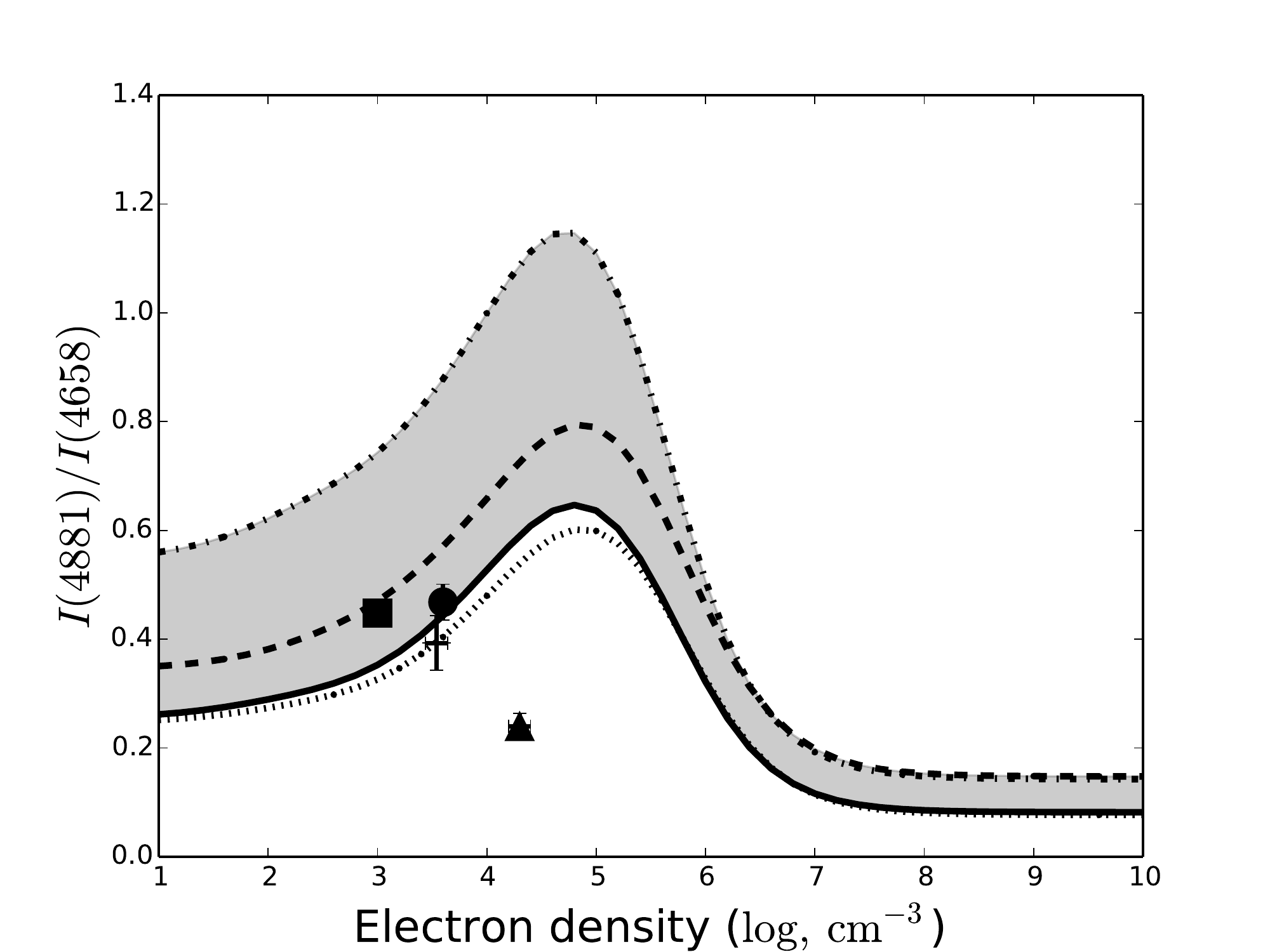} 

\caption{ Same as Figure \ref{fig:1} except for the ratio 4881/4658.}\label{fig:5}
\end{figure*}

\begin{figure*}
  \centering

\includegraphics[width=10cm,angle=0]{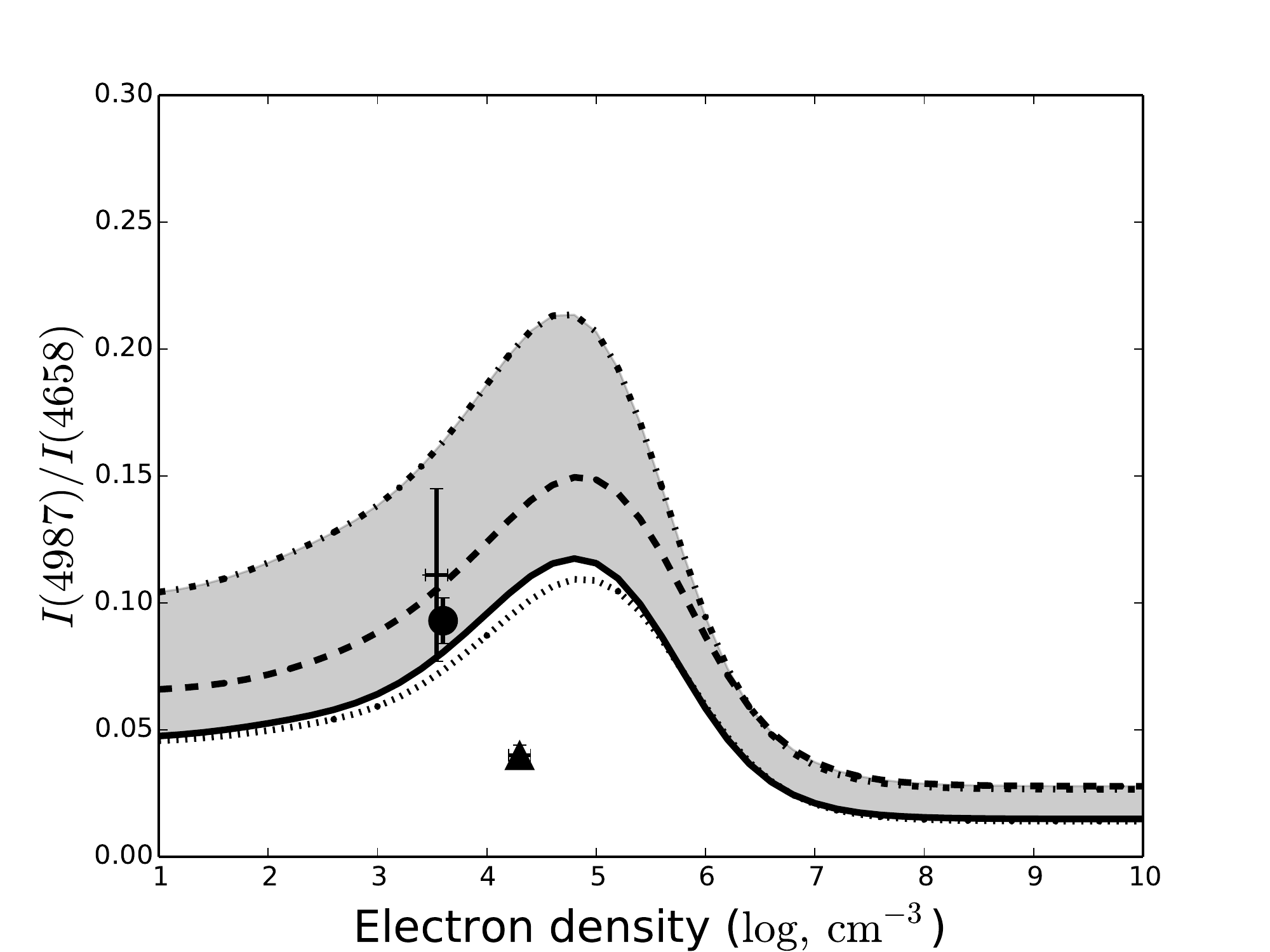}

\caption{Same as Figure \ref{fig:1} except for the ratio 4987/4658. Note that the triangle is an upper limit to the 
observed ratio ($\leq$0.04).}\label{fig:6}
\end{figure*}

\begin{figure*}
  \centering 

\includegraphics[width=10cm,angle=0]{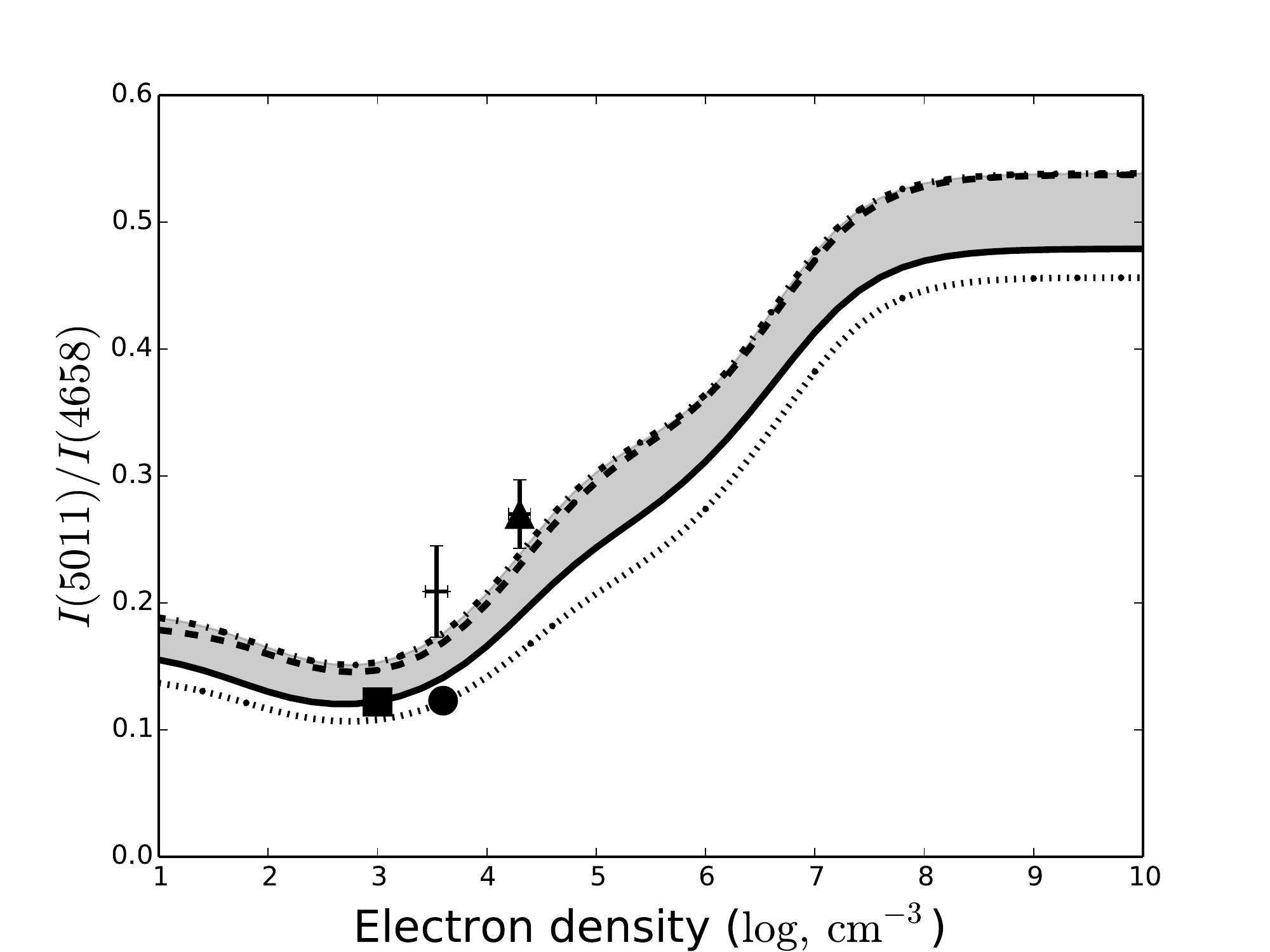}

\caption{Same as Figure \ref{fig:1} except for the ratio 5011/4658.}\label{fig:7}
\end{figure*}

\begin{figure*}
  \centering

\includegraphics[width=10cm,angle=0]{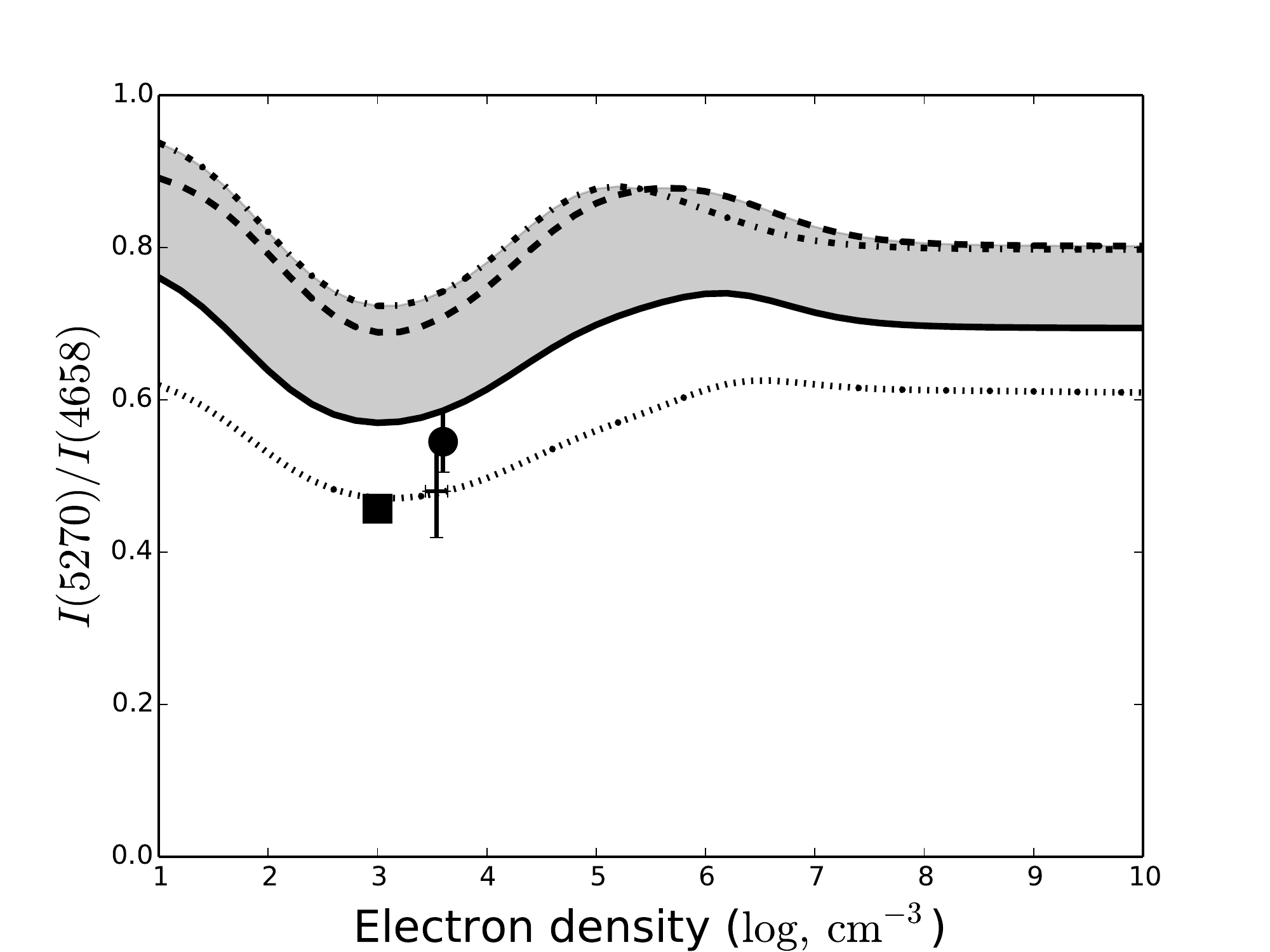} 

\caption{Same as Figure \ref{fig:1} except for the ratio 5270/4658}\label{fig:8}
\end{figure*}

\begin{figure}[h!]
  \centering 
\includegraphics[width=10cm,angle=0]{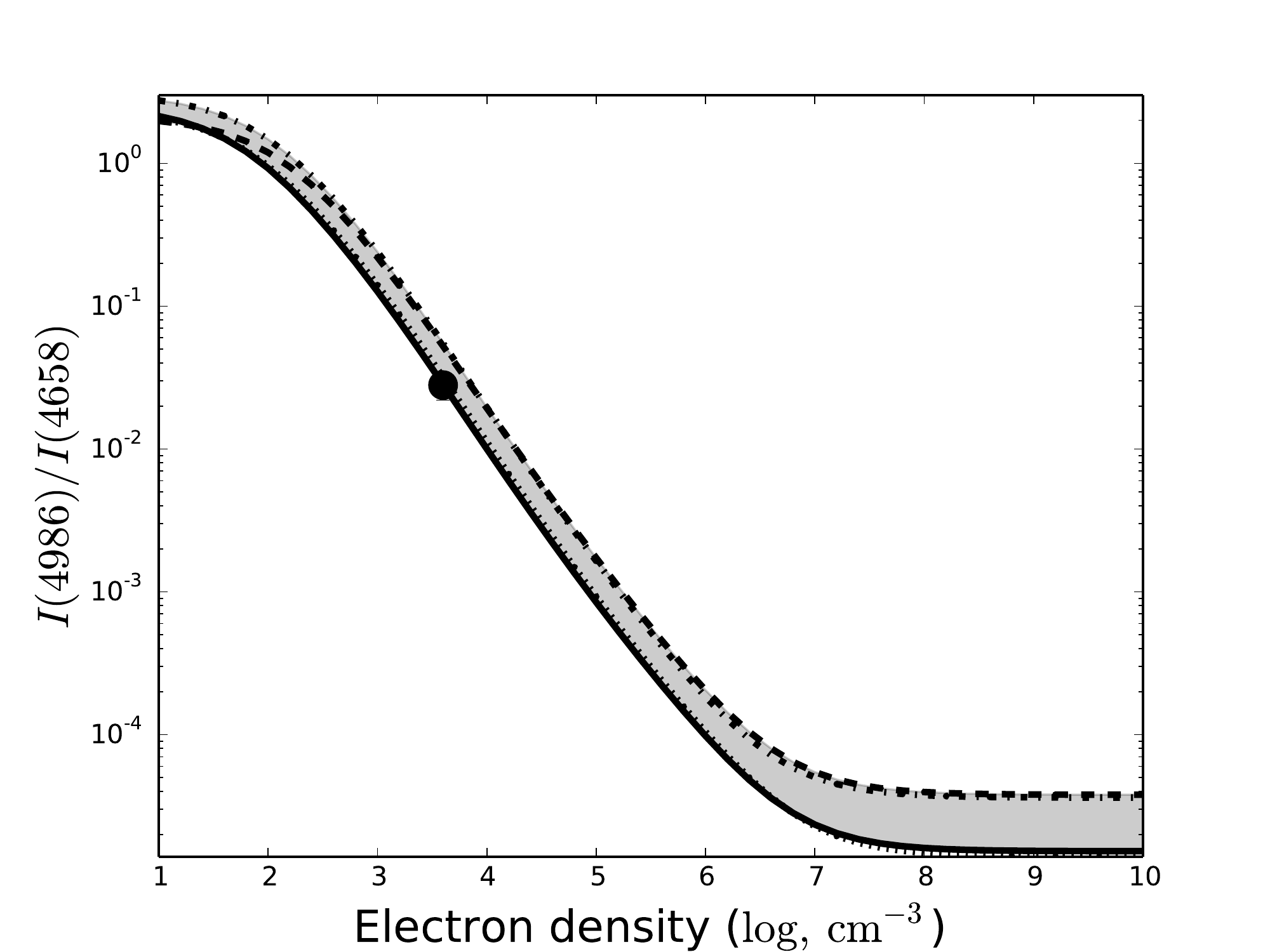} 
\caption{Same as Figure \ref{fig:1} except for the ratio 4986/4658.  }\label{fig:9}
\end{figure}

Figures \ref{fig:1}--\ref{fig:9} show a number of density-sensitive \fe emission line ratios, generated 
using the three Cloudy models CLOUDY1, CLOUDY2 and CLOUDY3 as a function of $N_{\rm e}$ in the range 10 -- $10^{10} \cmcubei$. In our 
calculations we have adopted a temperature of $9000$ K, to match those found for the observed nebulae in Table \ref{Table:obs3}. However, to show the 
temperature dependence of the ratios, we also plot results at $T_{\rm e}$ = 15000 K in Figures \ref{fig:1}--\ref{fig:9}. We find that
apart from 5011/4658 and 5270/4658 in Figures \ref{fig:7} and \ref{fig:8}, all the ratios are relatively insensitive to temperature variations.

The observed values of the \fe line ratios are plotted in Figures \ref{fig:1}--\ref{fig:9} at the electron densities listed for 
the source in Table 4. Also indicated in each figure is the range in the theoretical line ratios from the various Cloudy models, 
which arises mainly due to the adoption of different sets of A-values in each. Note that 
in all cases we use the ECS data of \citet{2014ApJ...785...99B}, although the different energy level values in various models
will result in somewhat different excitation and de-excitation rates, and may hence impact the line ratios. 
The spread in ratio values may be interpreted as `error bands' in the calculations. For most ratios, this error band
is 20 -- 30\% of the CLOUDY3 curve values. However, for 4881/4658 and 4987/4658 in Figures \ref{fig:5} and \ref{fig:6}, the error bands are more than 50\% of the 
CLOUDY3 ratio values, indicating large differences in the A-values for these transitions in the Cloudy models.

 We have also estimated the errors arising in the line ratios due to 
 possible uncertainties in the ECS calculated by \citet{2014ApJ...785...99B}. The resonances in the electron scattering cross section near threshold may sometimes have high peaks, and yield higher values of the ECS. 
 We have removed those values of cross section which have only one point above the local average, and then convolved the 
 remainder with Gaussian profiles with full-width-half-maxima of 40 meV. The resulting ECS differ by less than
 1\% from the original values, which in turn does not therefore modify the theoretical line ratios in any model. {From Table 2 of \citet{2014ApJ...785...99B} we note that the ECS calculated by the authors using three different methods (ICFT, BPRM and DARC) agree very well. The differences in the values do not affect the line ratios calculated in this work.} Hence, we 
 only focus on the differences in the A-values as a possible source of errors for the \fe line ratios.

For line ratios having common upper levels, and which hence should be density and temperature independent, we have calculated  theoretical values at $N_e=10^4 \cmcubei$ and $T_e=9000$ K using the CLOUDY3 model. These are listed in Table \ref{Table:obs2}.

%%%%%%%%%%%%%%%   Section 4

\section{Results and discussion}\label{section:results}

\subsection{Emission line intensity ratios}

An inspection of Figures \ref{fig:1}--\ref{fig:9} reveal that the observed \fe line ratios 
mostly lie within the error bands of the theoretical values calculated at $T_{\rm e}$ = $9000$ K, except for 5270/4658. 
In particular, they are generally in best agreement with line ratios calculated with the CLOUDY3 model, which we believe contains the 
most reliable atomic dataset, although within the error bars in the observations, the results are consistent with all three models.
However, in the case of 5270/4658 in Figure 4, the measured ratios lie outside all of the Cloudy model ranges at $T_{\rm e}$ = 9000 K. As the other ratios do not show a significant temperature sensitivity, and hence the observations are in reasonable agreement 
with the $T_{\rm e}$ = 15000 K calculations as well as those at 9000 K, it is possible that the \fe - emitting region of the plasma 
is at a much higher temperature than indicated from other spectral diagnostics. We point that these diagnostics 
do indicate a range of temperatures and not a unique value. For example, \citet{1998MNRAS.295..401E} find $T_{\rm e}$ = 9000 -- 12400 K 
for the Orion nebula. However, it is difficult to believe that the temperature of the \fe region could be so different from
those of other ions. We note that there are also significant discrepancies between theory and observation for 4881/4658 and 4987/4658 in HD 341617, as previously noted by \citet{2003A&A...401.1119R}, although this is not the case for these ratios in the other sources. It is therefore possible that there is some error in the measurements of 4881/4658 and 4987/4658 in the Ryans et al. spectrum. 

To investigate if the discrepancies between theory and observation may be due to line blending, in Table \ref{Table:obs2}
we list measured line ratios involving transitions from common upper levels (which hence should be $T_{\rm e}$- and $N_{\rm e}$-independent), plus the calculated values from the CLOUDY3 model. However, we note that the theoretical results are similar from all three models.
An inspection of the table reveals good agreement between theory and observation, including for the ratio 
with the 4658\,\AA\ line, the transition in common for the $N_{\rm e}$-diagnostics. We can therefore rule out blending as a likely cause of the 
observed discrepancies. Hence, below we investigate if the atomic data may be responsible for these. 

Previous calculations of \fe line ratios have employed A-values and ECS from a variety of sources, with some difference from 
those presented here, but most in agreement. For example, \citet{1993ApJ...410..430K}, henceforth K93, 
 have derived the density dependent line ratios of \fe using A-values from \citet{1957MNRAS.117..393G} and ECS from \citet{1991JPhB...24.3467B}. A comparison of these with results from our CLOUDY3 model is shown in Table 6 at $T_{\rm e}=10,000$ K and $N_{\rm e}=10^4\cmcubei$. 
 We find that there are no major differences between our calculations and those of K93 with the exception of 
 5011/4658 and 5270/4658. The low density tail of the latter derived by K93 reaches a value of $\sim 0.3$, while with the latest atomic data the line ratio is mostly flat (with value of $\sim 0.7$) and insensitive to density. This is understandable from the fact that the A-value for the $5270 \rm \AA$ transition in \citet{1957MNRAS.117..393G} is 0.355 s$^{-1}$  while it is 0.570 s$^{-1}$ in \citet{2009ADNDT..95..184D}. Similarly for $5011 \rm \AA$ the A-value of \citet{1957MNRAS.117..393G} is 0.473 s$^{-1}$ and 
 0.770 s$^{-1}$ in \citet{2009ADNDT..95..184D}.
 
\begin{figure}[h!]
  \centering 

\includegraphics[width=10cm,angle=0]{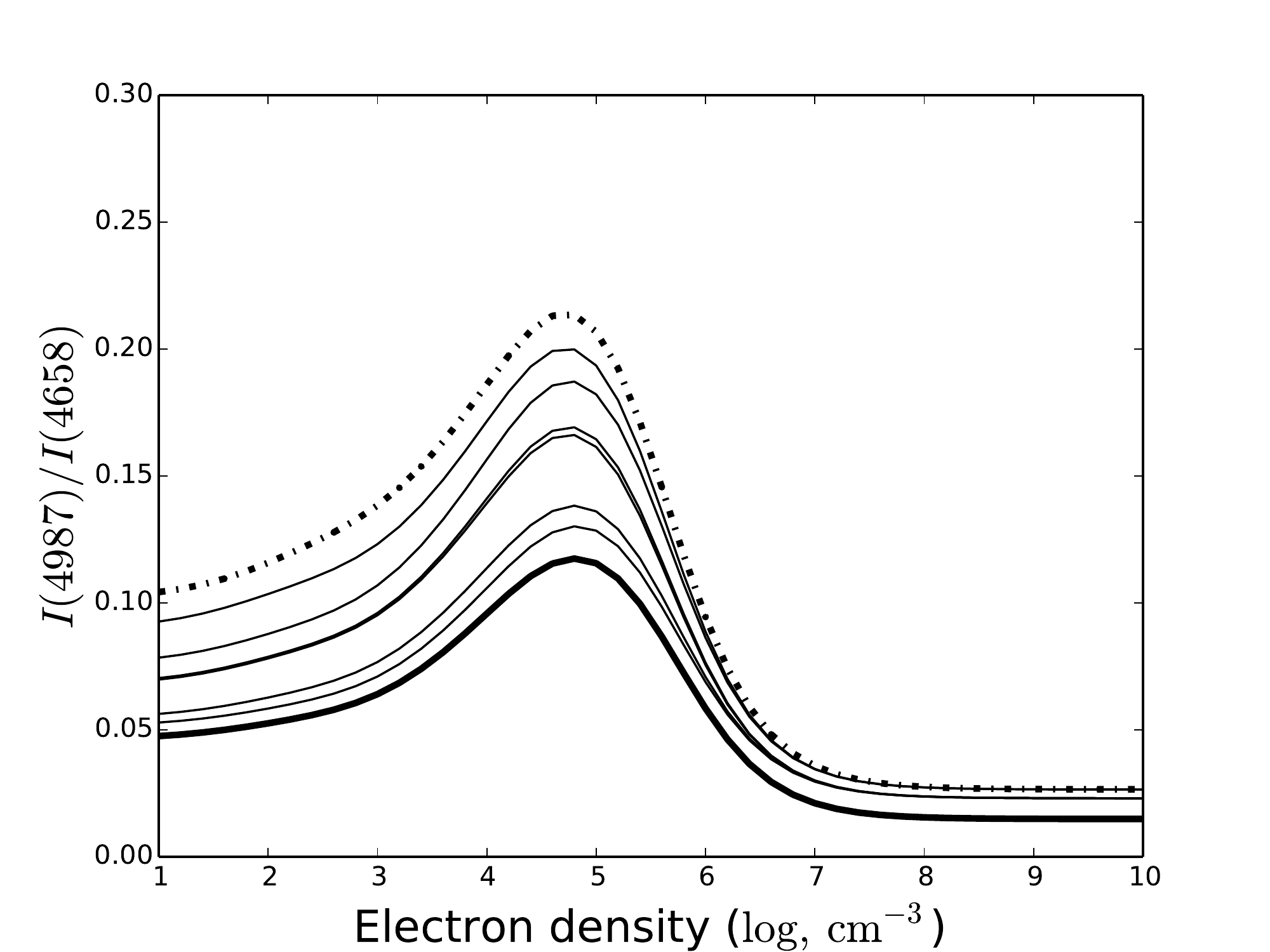} 
\caption{The line ratio $4987/4658$ plotted at $T_{\rm e}=9000$~K as a function of electron density. The top curve is obtained using the CLOUDY2 dataset while the bottom one is from CLOUDY3. Intermediate curves are obtained as we replace the A-values for the transitions involving the levels $\rm ^3F_4$ and $\rm ^3H_4$ in the CLOUDY3 model with those of CLOUDY2 as listed in Table \ref{Table:Avalues} . }\label{fig:10}
\end{figure}

As noted earlier, we consider the differences in the line ratios from the three Cloudy models as error bands arising due mostly
 to the various sets of A-values adopted. We find that not only do the differences in the A-values of the relevant 
 transitions affect the ratios, but also those of others due to cascading of electrons from higher levels. To demonstrate 
 this we consider the example of 4987/4658 which shows a large error band ($>$50\%) between the CLOUDY2 and CLOUDY3 models. 
 The 4987 and 4658 lines are due to the $\rm ^5D_3 (2) - {^3H_4} (9)$ and $\rm ^5D_4(1) - {^3F_4}(12)$ transitions, respectively, where the bracketed quantities are the level numbers (with the ground state being level 1). 
 Cascading to levels 9 and 12 from 15 ($\rm ^3G_5$), 16 ($\rm ^3G_4$) and 17 ($\rm ^3G_3$) is 
 important because of the relatively large transition probabilities, and also the fact that 
 15, 16 and 17 are closely spaced in energy (3.04, 3.09 and 3.11 eV, respectively). 
 In Table \ref{Table:Avalues} we list the A-values from the CLOUDY2 and CLOUDY3 models which involve pumping in and out of 
 levels 9 and 12, the upper levels of 4987 and 4658\,\AA, respectively. As we change each of the A-values from the CLOUDY3 to
 the CLOUDY2 data, the line ratio curves gradually move upwards as shown in Figure \ref{fig:10}. We find that cascading affects the \fe line ratios at a plasma temperature of $T_{\rm e}=9000$ K, and hence the corresponding A-values play an important role in deriving 
 the line ratios. However, in cases such as 4986/4658 in Figure \ref{fig:9}, where the error bar is small, 
 the line ratios may be effectively used to constrain the plasma density.  

\subsection{Plasma cooling function}

\begin{figure}[h!]
\centering 
\hbox{
\includegraphics[width=9cm,angle=0]{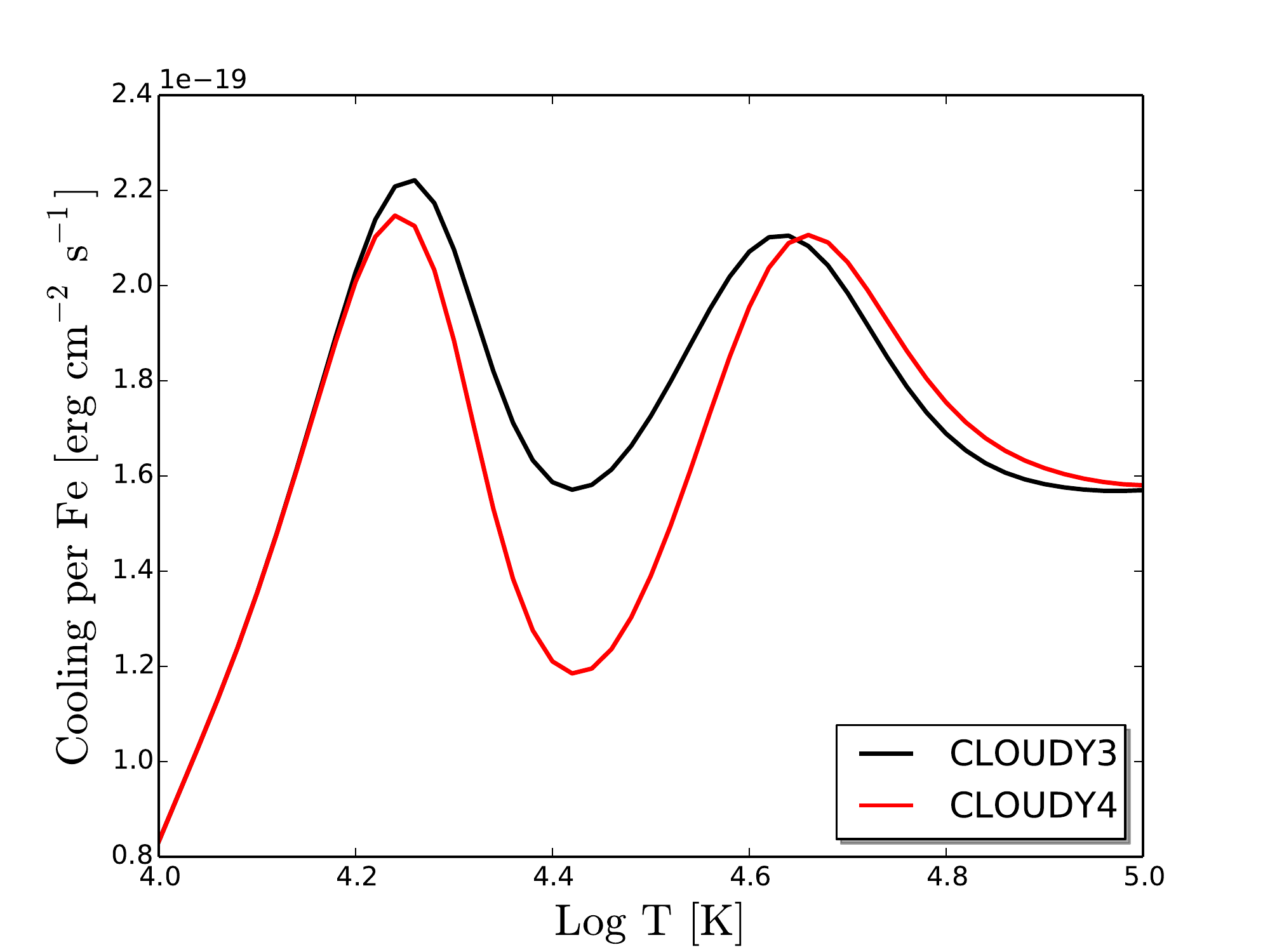} 
\includegraphics[width=9cm,angle=0]{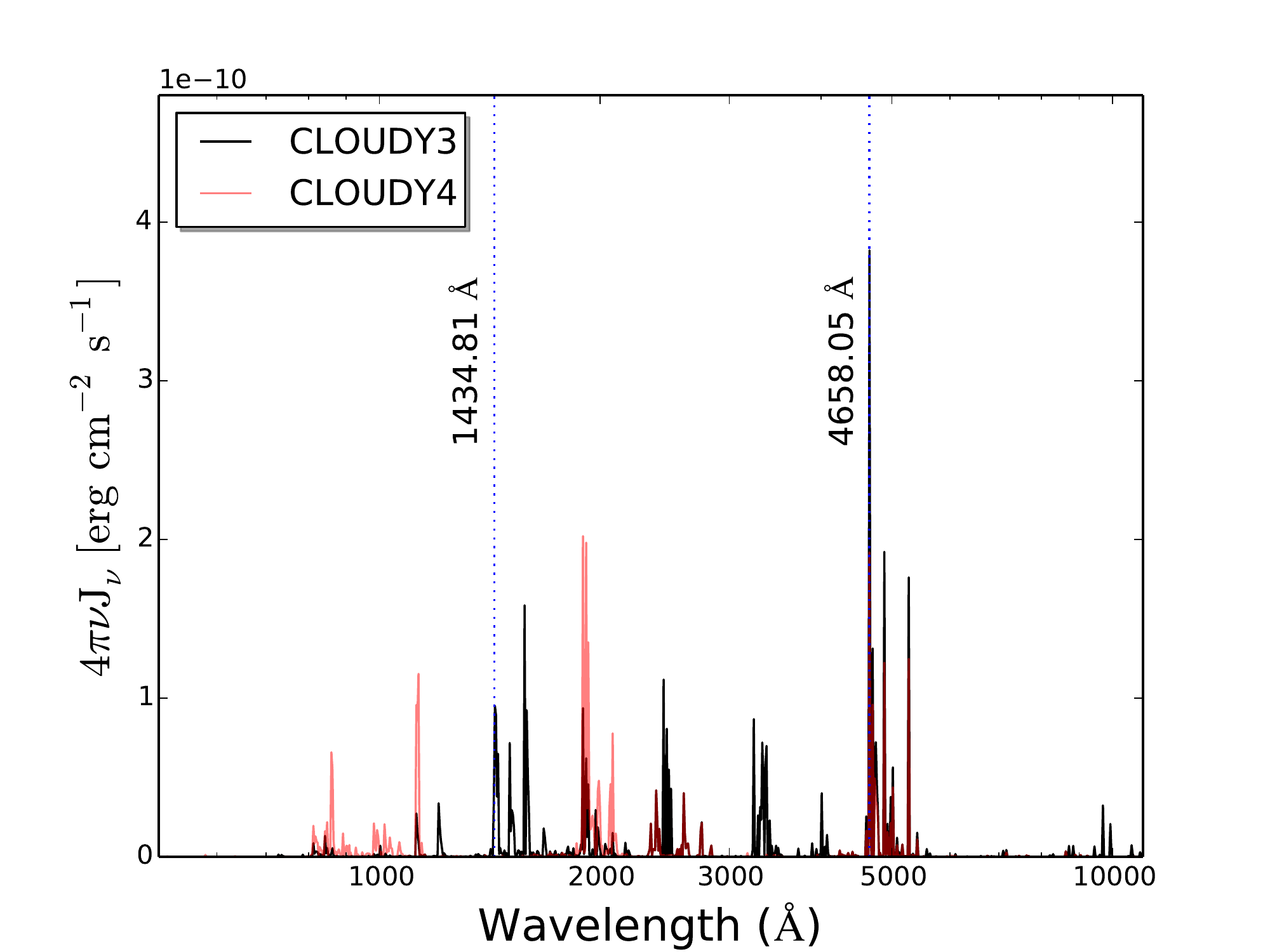} 
}\caption{{\it Left panel:} The total cooling function of a pure Fe plasma plotted as a function of electron temperature $T_{\rm e}$. The top curve is generated using the CLOUDY3 model, while the bottom one is from CLOUDY4. {\it Right panel:} Theoretical
 \fe spectra for a plasma at $T_{e}=30,000$ K and $N_{e}=1 \, \rm cm^{-3}$, generated using the CLOUDY3 and CLOUDY4 models.
See Section 4.2 for details.}\label{fig:11}
\end{figure}

We compare the total cooling function for a pure Fe plasma generated using two different atomic datasets, namely that
currently used in Cloudy (CLOUDY4) and the final model (CLOUDY3). Iron is known to be an important contributor to the cooling function
in nebulae, which is a fundamental parameter since it determines the thermal stability and energy balance of the plasma \citep{2012ApJS..199...20G}. We have repeated the cooling function calculations described by \citet{2013MNRAS.429.3133L} using 
both the CLOUDY3 and CLOUDY4 models, and in Figure \ref{fig:11} plot these for a temperature range over which
\fe has a significant fractional abundance. The cooling at $\simeq$ 30,000 K is enhanced when the new CLOUDY3 data are used,
which could have a major impact on the thermal stability of environments near this temperature. 
Figure \ref{fig:11} also shows a comparison of the \fe spectra predicted with the two datasets.  There are several 
regions where CLOUDY3 predicts lines while CLOUDY4 does not, with the largest difference for the UV/near-UV region between
1000--4000\,\AA. This is due to the larger number of levels in CLOUDY3 (322 with E$_{\rm max}$ = 221274 cm$^{-1}$ compared to 219 
with E$_{\rm max}$ = 137522 cm$^{-1}$ in CLOUDY4), combined with the 
inclusion of A-value data in this model which are absent in CLOUDY4. For example, consider one of the strongest lines in the 
spectrum which is present in CLOUDY3 but not in CLOUDY4, namely that  at 1434.81\,\AA\ (See Figure \ref{fig:11}). This
 line arises due to a transition from level 43 (E = 69695 cm$^{-1}$) to the ground state, and its A-value 
 in CLOUDY3 is 113.74 s$^{-1}$, whereas CLOUDY4 does not contain an A-value for this transition, explaining 
 its absence. The larger number of emission lines in CLOUDY3 in turn leads to 
 additional cooling, as indicated in Figure \ref{fig:11}.

\section{Conclusions}

We conclude that the existing discrepancies between theory and observation for \fe line ratios in nebular plasmas 
cannot be fully resolved using currently available atomic data. Furthermore, blending 
of the \fe lines is unlikely to be the cause because theoretical temperature- and density-independent \fe line ratio values involving
transitions from common upper levels are in agreement with measured values. However, we find that 
cascading of electrons from the $\rm 3d^6$ $\rm ^3G_5$, $\rm ^3G_4$ and $\rm ^3G_3$ levels plays an important role in populating the levels which provide the diagnostic emission lines of \fe. Hence, the A-values for these transitions are crucial in determining the line ratios, and further calculations for these would be highly desirable, as would measurements if feasible. We note that some of the \fe line ratios in Figures \ref{fig:1}--\ref{fig:9} do show good agreement between theory and measurement, including 4734/4658, 4778/4658, and 4986/4658, and hence may be employed as $N_{\rm e}$-diagnostics. The most reliable is probably 4986/4658 in Figure \ref{fig:9}, due to the narrow error band and lack of $T_{\rm e}$ sensitivity. However, the 4986\,\AA\ line is often weak and may not always be detected. 

Adoption of the most recent \fe atomic data in Cloudy leads to a cooling function which is significantly greater around 30,000\,K
than that generated with the existing Cloudy model. This is due to the presence of more emission lines in the former,
particularly in the UV/near-UV wavelength range from 1000--4000\,\AA.

%%%%%%%%%%%%%%%%%%%%%%%%%%%%%%%%%%%%%  Acknowledgements...

\section*{ACKNOWLEDGEMENTS}

 The project has made use of public databases hosted by SIMBAD, maintained by CDS, Strasbourg, France. SL, CAR and FPK are grateful to STFC for financial support via grant ST/L000709/1. GJF acknowledges financial support from the Leverhulme Trust via Visiting Professorship grant VP1-2012-025, and also support by the NSF (1108928, 1109061 and 1412155), NASA (10-ATP10-0053, 10-ADAP10-0073, NNX12AH73G and ATP13-0153) and STSciI (HST-AR-13245, GO-12560, HST-GO-12309, GO-13310.002-A, HST-AR-13914 and HST-AR-14286.001). CHIANTI is a collaborative project involving George Mason University, the University of Michigan (USA) and the University of Cambridge (UK). 
 SL is grateful to the CHIANTI Helpdesk Team and Peter Young in particular for their help with CHIANTI software.

%%%%%%%%%%%%%%%%%%%%%%%%%%%%%%     Figures..

%%%%%%%%%%%%%%%%%%%%%%%%%%%%%%  Fig  1.   

%%%%%%%%%%%%%%%%%%%%%%%%%%%%%   Fig  2.

%%%%%%%%%%%%%%%%%%%%%%%%%%%%%    Fig  3.  

%%%%%%%%%%%%%%%%%%%%%%%%%%%%    Fig  4.

%%%%%%%%%%%%%%%%%%%%%%%%%%%      Fig 5.

%%%%%%%%%%%%%%%%%%%%%%%%%%%%     Fig 6.

%%%%%%%%%%%%%%%%%%%%%%%%%%%%      Fig 7.

%%%%%%%%%%%%%%%%%%%%%%%%%%%     Fig  8.

%%%%%%%%%%%%%%%%%%%%%%%%%%%%%   Fig  9.  

%%%%%%%%%%%%%%%%%%%%%%%%%%%%     Fig.  10.  

%%%%%%%%%%%%%%%%%%%%%%%%%%   Fig 11.

%%%%%%%%%%%%%%%%%%%%%%%%%%%%%%%%%%%%  Table 1 ... the transitions

\newpage
\begin{table*}
%\begin{sidewaystable}
%\footnotesize
\centering
%\begin{minipage}{140mm}
\caption{  \fe emission lines.} \label{Table:atomic-data}
  \begin{tabular}{lccccccccccccccccc} \hline\hline 

  Lower level$\rm ^a$	& Upper level$\rm ^a$ 	& Wavelength ($\rm \AA$) & Wavelength ($\rm \AA$)$^b$	\\
			&			& (NIST, Vacuum) 	& (Adopted)	  \\  \hline 
$\rm ^5D_4$		&$\rm ^3G_4$	&  $\rm 4008.36$   &  $\rm 4008$   			&\\                                         
$\rm ^5D_3$		&$\rm ^3G_3$	&  $\rm 4046.43$   &  $\rm 4046$   			&\\                                         
$\rm ^5D_3$		&$\rm ^3G_4$	&  $\rm 4079.70$   &  $\rm 4080$   			&\\                                         
$\rm ^5D_2$		&$\rm ^3G_3$	&  $\rm 4096.61$   &  $\rm 4097$     		&\\ 
$\rm ^5D_4$		&$\rm ^3F_3$	&  $\rm 4607.03$   &  $\rm 4607$   			&\\     
$\rm ^5D_4$		&$\rm ^3F_4$	&  $\rm 4658.05$   &  $\rm 4658$   			&\\ 
$\rm ^5D_3$		&$\rm ^3F_2$	&  $\rm 4667.01$   &  $\rm 4667$     		&\\      
                                                                           
 $\rm ^5D_3$		&$\rm ^3F_3$	&  $\rm 4701.53$   &  $\rm 4702$   			&\\                                         
                                                                           
$\rm ^5D_2$		&$\rm ^3F_2$	&  $\rm 4733.91$   &  $\rm 4734$   			&\\
                                                                           
$\rm ^5D_3$		&$\rm ^3F_4$	&  $\rm 4754.69$   &  $\rm 4755$     		&\\                                                
	                                                                   
$\rm ^5D_2$		&$\rm ^3F_3$	&  $\rm 4769.43$   &  $\rm 4769$   			&\\                                         
	                                                                   
$\rm ^5D_1$		&$\rm ^3F_2$	&  $\rm 4777.68$   &  $\rm 4778$   			&\\                                         
	                                                                   
$\rm ^5D_4$		&$\rm ^3H_4$	&  $\rm 4881.00$   &  $\rm 4881$   			&\\ 
                                                                           
$\rm ^5D_4$		&$\rm ^3H_5$	&  $\rm 4924.50$   &  $\rm 4925$   			&\\  	
$\rm ^5D_1$		&$\rm ^3P_0$	&  $\rm 4930.54$   &  $\rm 4931$     		&\\                                                   
	                                                                   
$\rm ^5D_4$		&$\rm ^3H_6$	&  $\rm 4985.90$   &  $\rm 4986$   			&\\                                         
	                                                                   
$\rm ^5D_3$		&$\rm ^3H_4$	&  $\rm 4987.20$   &  $\rm 4987$   			&\\                                         
	                                                                   
$\rm ^5D_2$		&$\rm ^3P_1$	&  $\rm 5011.26$   &  $\rm 5011$   			&\\                                         
$\rm ^5D_0$		&$\rm ^3P_1$	&  $\rm 5084.77$   &  $\rm 5085$     		&\\                                         
$\rm ^5D_3$		&$\rm ^3P_2$	&  $\rm 5270.40$   &  $\rm 5270$     		&\\                                         
$\rm ^5D_1$		&$\rm ^3P_2$	&  $\rm 5411.98$   &  $\rm 5412$     		&\\                              	
                                        
\hline \hline
\end{tabular}
 
\begin{minipage}{6.5cm}%
{$\rm ^a$ All levels are within the $\rm 3d^6$ ground state configuration. } \\
{$\rm ^b$ Wavelengths used in this paper for brevity.}
\end{minipage}%

%\end{minipage}

%\end{sidewaystable}
\end{table*}

%%%%%%%%%%%%%%%%%%%%%%%%%%%%%%%%%%%%%%%%%%%%%%%%%%%%%%        Table 2...   Density dependent ratios  

%\begin{landscape}
\begin{table*}
%\begin{sidewaystable*}
{\footnotesize
\centering
\begin{minipage}{10cm}
\centering
\caption{ Observed \fe line ratios predicted to be density dependent.} \label{Table:obs1}
  \begin{tabular}{lccccccccccccccccc} \hline\hline

Line ratio 	& { HD~341617}$^1$ 	&{HH~202 }$^2$ 				&{ Orion}$^3$			&Orion$^4$\hspace{0.5cm}			\\ \hline 
                                                                                                                                                         
4702/4658			&0.37			&$0.27 \pm 0.03$			&$0.31 \pm 0.02$		&$0.31 \pm 0.02$	        	\\
                                                                                                                                                         
4734/4658			&0.21			&$0.14 \pm 0.02$			&$0.13 \pm 0.01$		&$0.12 \pm 0.01$	        	\\
                                                                                                                                                         
4769/4658			&0.15			&$0.11 \pm 0.02$			&$0.11 \pm 0.01$		&$0.11 \pm 0.01$	        	\\
                                                                                                                                                         
4778/4658			&0.07			&$0.04 \pm 0.01$			&$0.06 \pm 0.01$		&$0.05 \pm 0.01$	        	\\
                                                                                                                                                         
4881/4658			&0.24			&$0.39 \pm 0.05$			&$0.45 \pm 0.01$		&$0.47 \pm 0.03$	        	\\
                                                                                                                                                         
4986/4658			&-			&-					&-				&$0.03 \pm 0.01$	        	\\
                                                                                                                                                         
4987/4658			& $\leq$0.04		&$0.11 \pm 0.03$			&-				&$0.09 \pm 0.01$	 		\\
                                                                                                                                                         
5011/4658			&0.27			&$0.21 \pm 0.04$			&$0.12 \pm 0.01$		&$0.12 \pm 0.01$	        	\\
                                                                                                                                                         
5270/4658			&-			&$0.48 \pm 0.06$			&$0.46 \pm 0.01$		&$0.54 \pm 0.04$	        	\\

\hline \hline
\end{tabular} 
\end{minipage}%

\begin{minipage}{8cm}
\begin{flushleft}
{$^1$ \citet{2003A&A...401.1119R}. } \\
{$^2$ \citet{2009MNRAS.395..855M}. } \\
{$^3$ \citet{2000ApJS..129..229B}. } \\
{$^4$ \citet{1998MNRAS.295..401E}. } \\
\end{flushleft}
\end{minipage}%

}
%\end{sidewaystable*}
\end{table*}
%\end{landscape}

%%%%%%%%%%%%%%%%%%%%%%%%%%%%%%%%%%%%%%%%%%%%%        Table 3...   Density independent ratios

\begin{table*}
%\begin{sidewaystable}
\centering

{\footnotesize
\begin{minipage}{9cm}
\centering
\caption{ \fe line ratios having common upper levels.} \label{Table:obs2}
  \begin{tabular}{lccccccccccccccccccc} \hline\hline 
	
 Line ratio  & HD~341617$^1$	&HH~202$^2$ &Orion$^3$ & Orion$^4$	&CLOUDY3$^{5}$		\\ \hline

 4769/4702		&0.41	&$0.41 \pm 0.06$		&$0.34 \pm 0.03$			&$0.35 \pm 0.02$	&$0.34 $						\\  
 4778/4734		&0.33	&$0.29 \pm 0.08$		&$0.48 \pm 0.04$			&$0.42 \pm 0.03$	&$0.49 $						\\	
 4607/4702		&	&$0.27 \pm 0.05$		&$0.25 \pm 0.02$			&$0.26 \pm 0.01$	&$0.20$						\\	
                                                                                                                                                                          
 4667/4734		&-	&$0.38 \pm 0.12$		&$0.31 \pm 0.03$			&$0.51 \pm 0.04$	&$0.29$						\\
                                                                                                                                                                          
 4987/4881		& $\leq$0.17 & $0.28 \pm 0.09$  		& - 				&$0.20 \pm 0.01$	&$0.18$								\\  
                                                                                                                                                                          
 5085/5011		&-	&-				&$0.19 \pm 0.01$			&$0.22 \pm 0.02$	&$0.17$			        		\\	
 4756/4658		&-	&$0.19 \pm 0.03$		&$0.18 \pm 0.02$			&$0.18 \pm 0.01$	&$0.19$						\\

 4080/4008		&-	&-				&$0.28 \pm 0.04$			&-			&$0.28$						\\

 5412/5270		&-	&$0.13 \pm 0.05$		&$0.10 \pm 0.01$			&$0.08 \pm 0.01$	&$0.09$						\\

\hline \hline
\end{tabular} 
\end{minipage}

\begin{minipage}{9cm}
\begin{flushleft}
{$^1$ \citet{2003A&A...401.1119R}. } \\
{$^2$ \citet{2009MNRAS.395..855M}. } \\
{$^3$ \citet{2000ApJS..129..229B}. } \\
{$^4$ \citet{1998MNRAS.295..401E}. } \\
{$^{5}$ Line ratios calculated using the CLOUDY3 model for a plasma with 
$N_{\rm e}=10^4 \cmcubei$ and $T_{\rm e}=9000$ K.}
\end{flushleft}
\end{minipage}

}

%\end{sidewaystable}
\end{table*}

%%%%%%%%%%%%%%%%%%%%%%%%%%%%%%%%%%%%%%%%%  Table 4...    Average plasma properties..

\begin{table*}
%\begin{sidewaystable}

{\footnotesize
\centering
\begin{minipage}{14cm}
\centering
\caption{ Average plasma parameters derived from other emission line ratios.} \label{Table:obs3}
  \begin{tabular}{lcccccccccccccccccccc} \hline\hline 
	
Plasma					& HD~341617$^1$		&HH~202$^2$ 	&Orion$^3$  &Orion$^4$  		&\\ 
parameter				& \\  \hline

$\log(N_e \, , \, \cmcubei$)	&$4.3$			&$3.5\pm 0.09$		&$3.6\pm 0.1$	&$3.6\pm 0.1$			\\ \\ \hline \\

$(T_e \rm \, , \, K)$	& $ 10000^a$ 	&$ 9000$ 	&$ 9000$ 	&$9000$ 			\\
			
\hline \hline
\end{tabular} 

\begin{minipage}{8cm}%
\begin{flushleft}
{$^1$ \citet{2003A&A...401.1119R}. } \\
{$^2$ \citet{2009MNRAS.395..855M}. } \\
{$^3$ \citet{1991ApJ...374..580B}. } \\
{$^4$ \citet{1998MNRAS.295..401E}. } \\
{$^a$ Temperature obtained from \citet{2000A&A...357..241P}.} \\
\end{flushleft}
\end{minipage}

\end{minipage}
}
%\end{sidewaystable}
\end{table*}

%%%%%%%%%%%%%%%%%%%%%%%%%%%%%%%%%%%%  Table 5 ... the A values
\begin{table}
%\begin{sidewaystable}
{\footnotesize
%\centering
%\begin{minipage}{140mm}
\caption{ \fe A-values involving levels $\rm 3d^6$ $\rm ^3F_4$ and $\rm ^3H_4$} \label{Table:Avalues}
  \begin{tabular}{lccccccccccccccccccc} \hline\hline 

  Lower level	& Upper level 	& A-value ($\rm s^{-1}$)	&A-value ($\rm s^{-1}$) 	\\
		&			& CLOUDY3		 &CLOUDY2			\\ 

		&		&Deb \& Hibbert (2009)		&Badnell \& Ballance (2014)$^1$ 	\\ \hline
 
	$\rm ^5D_4$(1)		&$\rm ^3F_4$(12)	&0.5681				&0.3671				\\ 
	$\rm ^3F_4$(12)		&$\rm ^3G_5$(15)	&0.0182				&0.0074				\\ 
	$\rm ^3F_4$(12)		&$\rm ^3G_4$(16)	&0.0343				&0.0045				\\ 
	$\rm ^3F_4$(12)		&$\rm ^3G_3$(17)	&0.0002				&0.0071				\\

\\ \hline \\

	$\rm ^5D_3$(2)		&$\rm ^3H_4$(9)		&0.0077				&0.0089				\\ 
	$\rm ^3H_4$(9)		&$\rm ^3G_5$(15)	&0.0182				&0.0074				\\ 
	$\rm ^3H_4$(9)		&$\rm ^3G_4$(16)	&0.0343				&0.0045				\\ 
	$\rm ^3H_4$(9)		&$\rm ^3G_3$(17)	&0.0002				&0.0071				\\

\hline \hline
\end{tabular} 

{$^1$ The A-values from \citet{2014ApJ...785...99B} have been corrected for the energy differences with NIST. }

%\end{minipage}
}

%\end{sidewaystable}
\end{table}

%%%%%%%%%%%%%%%%%%%%%%%%%%%%%%%%%%%%%%%%%%%%%%%%   Table 6 ... Comparison with K93.

\begin{table}
%\begin{sidewaystable}

{\footnotesize
\centering
%\begin{minipage}{140mm}
\caption{ Comparison of theoretical \fe line ratios with previous work. } \label{Table:K93}
  \begin{tabular}{lccccccccccccccccccc} \hline\hline 
	
Line Ratio \hspace{0.6cm}	& Keenan et al. (1993)	&CLOUDY3$^1$ \\ \hline 

$4702/4658$		&0.34		&0.33\\	
$4734/4658$		&0.14		&0.14\\	

$4778/4658$		&0.06		&0.06	\\	
$4881/4658$		&0.38		&0.51	\\	
$5011/4658$		&0.07		&0.16	\\	
$5270/4658$		&0.32		&0.59	\\

\hline \hline
\end{tabular} 

{$^1$Calculated at $T_e=10^4$ K and $N_e=10^4 \cmcubei$.}

%\end{minipage}
}
%\end{sidewaystable}
\end{table}

\bibliographystyle{apj}
\bibliography{mybib}

\end{document}